\documentclass[aps,prl,reprint,groupedaddress,showpacs]{revtex4-1}
\usepackage{graphics} % or graphicx
\usepackage{amsmath}
\usepackage{amsfonts}
\usepackage{amssymb}
\usepackage{bm}

\begin{document}

\title{Extending the Electron Spin Coherence Time of Atomic Hydrogen by Dynamical Decoupling}

\author {George Mitrikas}
\email[Electronic address: ]{mitrikas@ims.demokritos.gr}
\author {Eleni K. Efthimiadou}
\author {George Kordas}

\affiliation{Institute of Advanced Materials, Physicochemical Processes, Nanotechnology and Microsystems, NCSR Demokritos, 15310 Athens, Greece}

\date{\today}

\begin{abstract}

We study the electron spin decoherence of encapsulated atomic hydrogen in octasilsesquioxane cages induced by the $^{1}$H and $^{29}$Si nuclear spin bath. By applying the Carr-Purcell-Meiboom-Gill (CPMG) pulse sequence we significantly suppress the low-frequency noise due to nuclear spin flip-flops up to the point where a maximum $T_2 = $ 56 $\mu$s is observed. Moreover, dynamical decoupling with the CPMG sequence reveals the existence of two sources of high-frequency noise: first, a fluctuating magnetic field with the proton Larmor frequency, equivalent to classical magnetic field noise imposed by the $^{1}$H nuclear spins of the cage organic substituents, and second, decoherence due to entanglement between the electron and the inner $^{29}$Si nuclear spin of the cage.

\end{abstract}

\maketitle

Hybrid electron-nuclear spin systems and in particular paramagnetic atoms trapped in molecular cages like endohedral fullerenes (e.g. N@C$_{60}$ or P@C$_{60}$) are promising components of spin-based quantum computing because they have long spin relaxation times and can be precisely placed into large arrays by chemical engineering \cite{Morton_review2011,Harneit2002}. Examples of recent progress include storage of quantum states over 100 ms \cite{Brown2011}, as well as design or preparation of new molecular architectures for scalable quantum computing \cite{Suter2011,Porfyrakis2012,Du2012}. Atomic hydrogen can be even more attractive due to its simpler electronic 1s state and the exceptionally large hyperfine coupling of 1420.406 MHz. Systems with large hyperfine couplings can under certain conditions benefit from fast manipulation of nuclear spins as was recently demonstrated in bismuth-doped silicon \cite{Morley_NatMat_2013}.

Whilst C$_{60}$ cannot stably host atomic hydrogen, it has been found that polyhedral octa-silsesquioxanes (POSS) are ideal cages for this purpose \cite{Matsuda1994}. Crucial properties for quantum computing like the spin-lattice $T_1$ and spin-spin $T_2$ relaxation times depend strongly on the type of the peripheral organic substituents \cite{Paech2006b}. Recently \cite{Mitrikas2012}, we showed that the room-temperature phase memory time $T_{\text{M}}$=13.9 $\mu$s for the species with R=OSiMe$_2$H is the longest observed so far for this kind of cages.

A key step in implementing a physical system as qubit is the identification of all possible sources of decoherence. Understanding the origin of magnetic interactions between the central spin and its environment can help to improve the performance of the qubit by prolonging the electron spin coherence using optimal dynamical decoupling \cite{Du_Nat_2009}. In addition, dynamical decoupling techniques allow for measuring noise spectral densities \cite{Yuge_PRL_2011,Suter_PRL_2011} and can thus map the electron spin environment. In this Letter we apply the CPMG pulse sequence on atomic hydrogen encapsulated in POSS cages (H@POSS) in order to suppress the low-frequency noise originating from the surrounding nuclear spin flip-flops. This process, also termed nuclear spin diffusion, is the main source of decoherence in paramagnetic systems containing magnetic nuclei, as is the present case of H@POSS. Therefore, the isolation of this decoherence mechanism can reveal the intrinsic spin-spin relaxation time which has to be compared with $T_2 = $160 $\mu$s, the longest coherence time of any solid state molecular radical that was recently reported for the $^{15}$N@C$_{60}$ system magnetically diluted in C$_{60}$ \cite{Brown2011}. Moreover, by applying dynamical decoupling we identify the nuclear spin baths surrounding the unpaired electron spin and we probe decoherence processes that were previously unexplored in paramagnetic trapped atoms.

\begin{figure}
%\centering
  \includegraphics{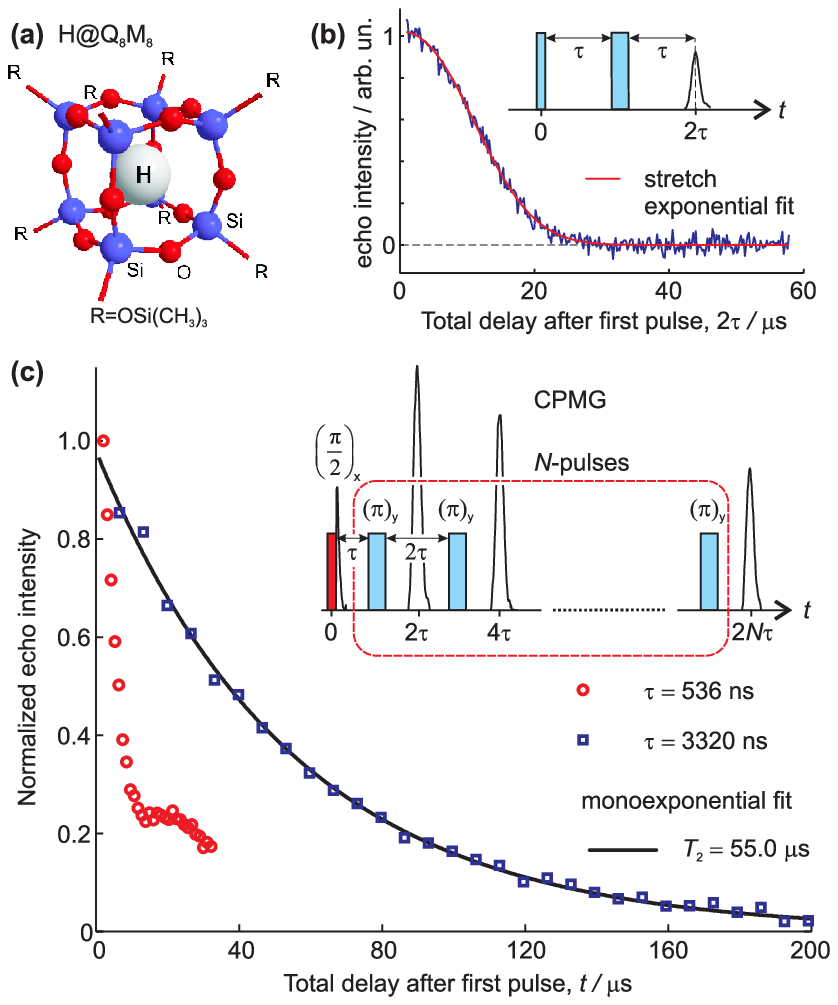}
  \caption{(Color online) (a) Structure of the POSS molecule Si$_8$O$_{12}$(OSiMe$_3$)$_8$ with the encaged H atom. (b) Two-pulse electron spin echo decay measured at the low-field EPR transition, $B_0=$318.6 mT, with superimposed stretched exponential fit using Eq. (\ref{eqn1}). (c) CPMG decay measured at $B_0=$318.6 mT with two different $\tau$ values and $N=30$ pulses as a function of the time delay between the first $\pi/2$ pulse and the occurring echoes, $t=2\tau, 4\tau,...,2N\tau$. Inset: CPMG sequence used for dynamical decoupling.}
  \label{fig1}
\end{figure}

Although CPMG sequence is a well-established and frequently used method in NMR spectroscopy \cite{Carr_Purcell_1954,Meiboom_Gill_1958}, its systematic application in EPR was only recently revisited as a dynamical decoupling method in the context of  spin-based quantum computing \cite{Du_Nat_2009,Wrachtrup_PRB_2011}. This is partly because paramagnetic systems have often broad EPR spectra that cannot be fully excited by microwave pulses leading to peculiarities \cite{Harbridge_2003}. Atomic hydrogen is an ideal testbed for applying CPMG decoupling because the solid state EPR spectrum exhibits narrow lines of the order of 0.1 mT.
In order to ensure a small delocalization of electron spin density over the cage, we used the species with R=OSiMe$_3$ shown in Fig.\ref{fig1}(a), also known as Q$_8$M$_8$ \cite{Matsuda1994}. Our previous study on the similar system with R=OSiMe$_2$H revealed that for temperatures below 200 K the spin-spin relaxation is strongly enhanced by dynamic processes like rotation of the methyl groups \cite{Mitrikas2012}. Therefore, in our effort to avoid this additional decoherence process, we chose to perform decoupling experiments at $T=$200 K where the upper limit of $T_2$ is virtually imposed by the spin-lattice relaxation time $T_1=$133 $\mu$s. Fig.\ref{fig1}(b) shows that at this temperature the two-pulse echo decay is described by a stretch exponential function
\begin{equation}
  E(2\tau)=E_0\,\mathrm{exp}\left[- \left (\frac{2\tau}{T_{\mathrm{M}}}\right)^n\right],
  \label{eqn1}
\end{equation}
where $\tau$ is the interpulse delay, $n$ is a parameter determined by the mechanism of phase memory decay and the rate, $W$, of the dephasing process relative to $\tau$, and $T_{\mathrm{M}}$ is the so-called phase memory time encompassing $T_2$ and all other processes that cause electron spin dephasing \cite{Eaton2002}. At $T=$200 K we find $T_{\mathrm{M}}$=14.4 $\mu$s and $n=$2.3, which implies a slow relaxation processes ($W\tau\ll1$) compatible with  nuclear spin diffusion \cite{Brown}.

Fig.\ref{fig1}(c) demonstrates that the CPMG decay measured with $\tau=$ 3320 ns (blue squares) is accurately described by a monoexponential function with time constant $T_2=55.0 $ $\mu$s. This is about four times longer than the value obtained with the two-pulse sequence, indicating a significant suppression of nuclear spin diffusion. Such dynamical decoupling is possible because the rate $W$ of nuclear spin flips is of the order of their dipolar interaction (typically $\sim$ 10 kHz for protons) corresponding to a correlation time $\tau_c\sim$16 $\mu$s which is much larger than the used $\tau$ value. Therefore, for short interpulse delays the decoherence process behaves like time-independent. Based on this concept, it is natural to assume that the longest relaxation decay corresponds to the minimum possible $\tau$ value whose lower bound is equal to the spectrometer deadtime, $\tau_d$. However, the CPMG decay obtained with $\tau=$ 536 ns (Fig.\ref{fig1} (c) red circles) shows clearly two peculiarities: first, the initial decay is surprisingly fast, even faster than the two-pulse echo decay; second, the total decay is not monotonic and cannot be described by a general stretch exponential. To the best of our knowledge, the dependence of the CPMG time constant on the time between the refocusing pulses has been documented once for an organic radical diluted in a mixture of deuterated solvents \cite{Harbridge_2003}, and has been correlated to the electron spin echo envelope modulation (ESEEM) effect \cite{Schweiger2001}. Moreover, by using CPMG pulse sequences on the nitrogen-vacancy center an anomalous decoherence effect has been recently reported \cite{Du_NatCom_2011}.

\begin{figure}
%\centering
  \includegraphics{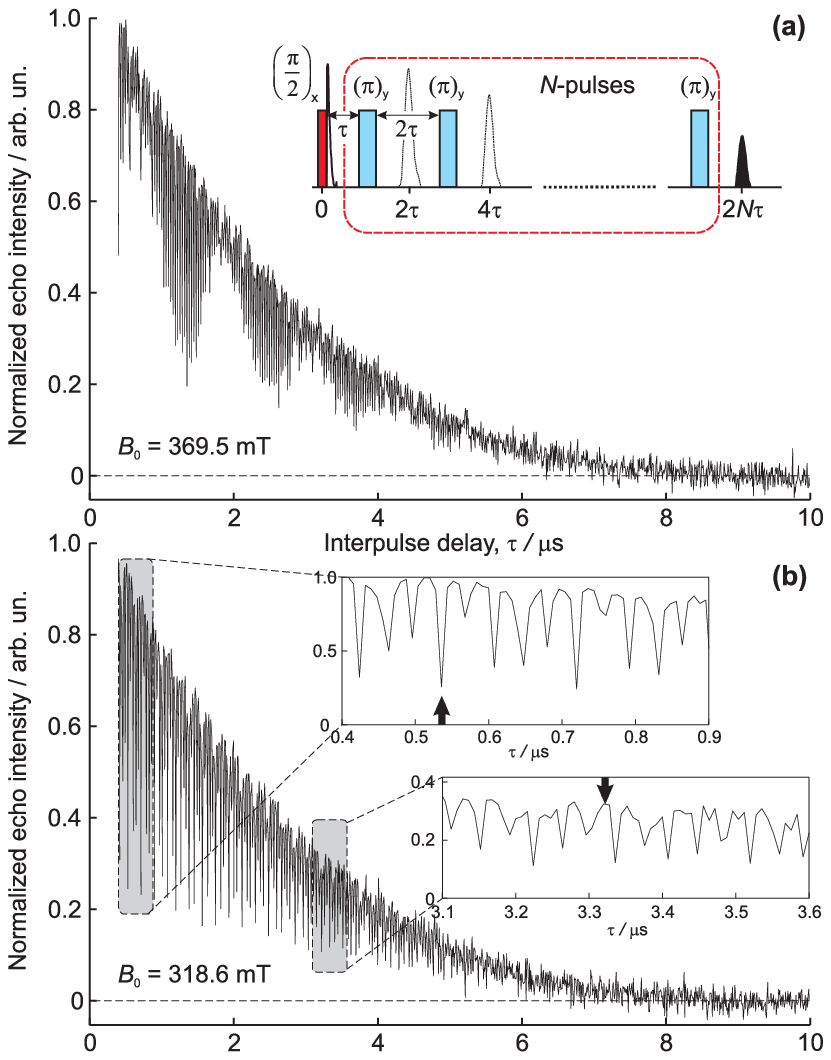}
  \caption{(Color online) Decay of the $N$'th echo in a CPMG sequence using $N=10$ as a function of interpulse delay $\tau$. (a) High-field EPR transition. Inset: pulse sequence showing the recorded echo. (b) Low-field EPR transition. The arrows in the enlarged time windows show the two $\tau$ values used in the measurements of Fig.\ref{fig1}.}
  \label{fig2}
\end{figure}

\begin{figure*}[t]
%\centering
  \includegraphics{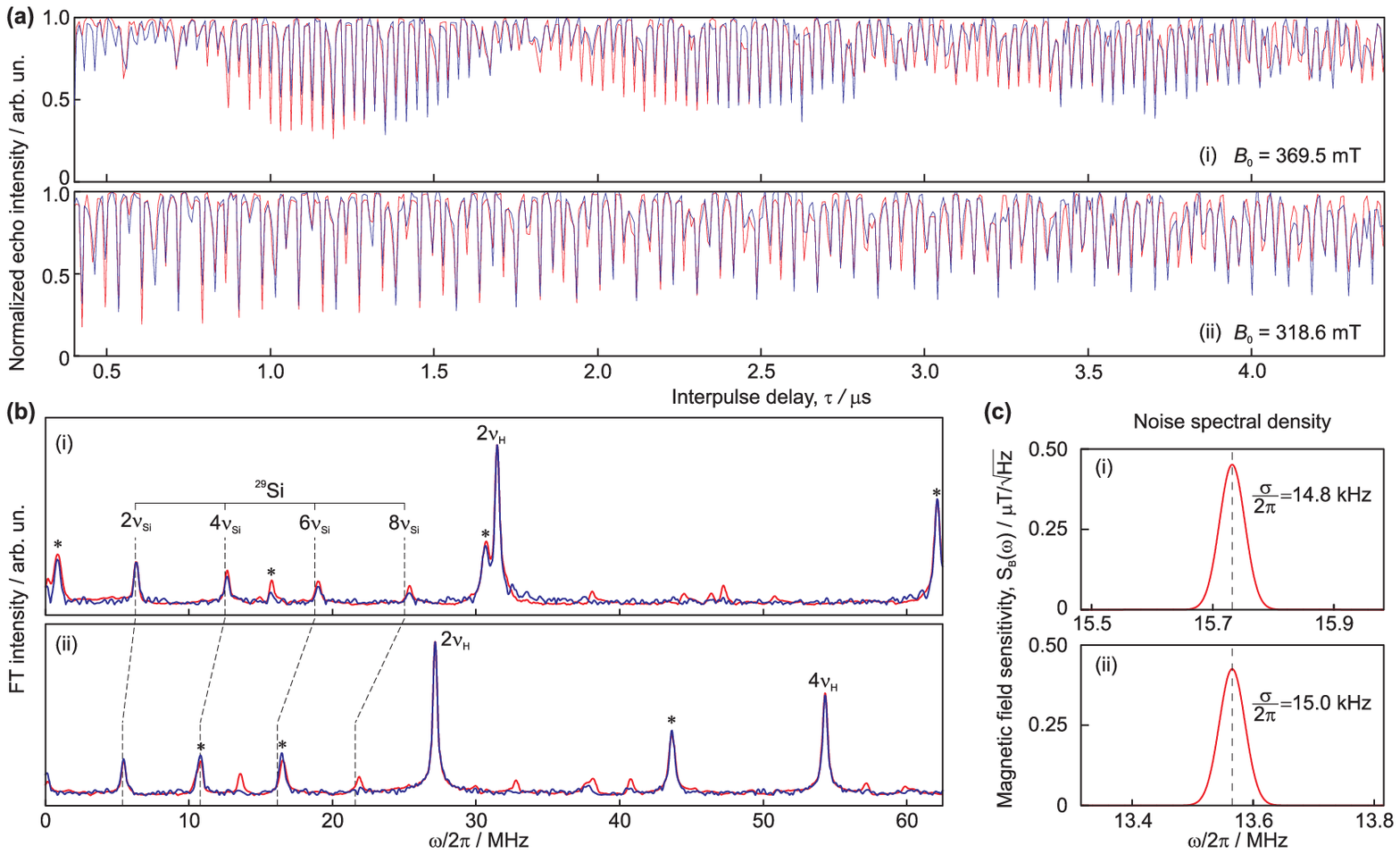}
  \caption{(Color online) (a) Baseline-corrected CPMG decays obtained with $N=$10 (blue traces) and their numerical simulations (red traces) for the two EPR transitions. (b) Absolute-value spectra obtained after FT of the time-domain experimental (blue lines) and simulated (red lines) signals. Asterisks mark "folded-back" peaks corresponding to even harmonics of the proton nuclear Zeeman frequency $\nu_H$ that are larger than the Nyquist frequency. (c) Noise spectral densities occurred from numerical simulations. Dashed lines denote the $^1$H Larmor frequency.}
  \label{fig3}
\end{figure*}

The above unusual behavior can be explained if we consider both the filter function of the applied CPMG pulse sequence, $\tilde{f}(t;\omega)=\int f(t,t')e^{-i\omega t'}dt'$, and the noise spectrum $S(\omega)$ of the system under study \cite{sup}. The decoherence rate is determined by the overlap of these two functions. For a minimum overlap the coherence is preserved for a long time, whereas for a maximum overlap the fastest decay occurs. Compared to the two-pulse sequence the filter function of the CPMG exhibits peaks at higher frequencies. Therefore, apart from suppressing low-frequency processes, it can also probe an existing high-frequency noise that is otherwise 'silent' in a Hahn echo experiment.

In order to understand the details of this peculiarity we performed CPMG experiments by measuring the intensity of the last echo of the pulse train as a function of $\tau$. This can give further insight into the spin dynamics of our system because each $\tau$ value corresponds to a different filter function of the CPMG sequence. Fig.\ref{fig2} demonstrates that in addition to the exponential decay the echo shows sharp dips with a modulation depth that exceeds 0.5 for some $\tau$ values. Applying dynamical decoupling with these $\tau$ values (e.g. $\tau$=536 ns as indicated by the up arrow of Fig.\ref{fig2}(b)) result in very fast CPMG decays (see red circles in Fig.\ref{fig1}) whereas measurements at maximum echo positions give exponential CPMG decays with time constants close to 60 $\mu$s.

The origin of the deep echo modulations can be best explored by analyzing the Fourier transform (FT) spectra shown in Fig.\ref{fig3}(b) which contain peaks close to even harmonics of the $^1$H and $^{29}$Si nuclear Zeeman frequencies. These modulations are typical in ESEEM spectroscopy and are attributed to weak anisotropic hyperfine couplings between the unpaired electron and matrix magnetic nuclei. Indeed, previous EPR studies have revealed such interactions both with protons of the cage organic substituents \cite{Stoesser1997} as well as with $^{29}$Si nuclei of the core POSS cage \cite{Mitrikas2012}. However, these couplings are weak and lead to shallow modulation depths as can be justified from the absence of dips in the two-pulse echo decay. Interestingly, the large number $N$ of refocusing pulses in the CPMG sequence enhances the modulation depth considerably. Similar behaviors have been recently observed on nitrogen-vacancy centers ($S=1$) \cite{Wrachtrup_NatNano_2012,Lukin_PRL_2012,Hanson_PRL_2012} where the CPMG sequence was used to detect very weak hyperfine couplings from remote $^{13}$C nuclear spins. In that case, and for large external magnetic fields (weak coupling regime, $\omega_I\gg B$), the basic modulation frequency is given by $\omega_{\pm}=|2\omega_I\pm A|$, where $\omega_I$ is the nuclear Zeeman frequency and $A, B$ are the secular and pseudosecular parts of the hyperfine interaction, respectively. This frequency is equal to the sum of $\omega_0=|\omega_I|$ and $\omega_{\pm1}=\sqrt{(\omega_I\pm A)^2+B^2}$, the corresponding ESEEM frequencies for $m_S=0$ and $m_S=\pm1$, in the limit of very small $B$ \cite{Schweiger2001}. For $S=1/2$, which is the case of our present study, the sum of the two ESEEM frequencies $\omega_{\alpha}, \omega_{\beta}$ is given by \cite{Schweiger2001}
\begin{equation}
  \omega_+=2|{\omega_I}|+\frac{|\omega_I|B^2}{4\omega_I^2-A^2}.
  \label{eqn2}
\end{equation}
Contrary to $S=1$, the weak coupling regime for $S=1/2$ results in sum frequencies that are virtually free from hyperfine coupling constants. This is the case for the weakly coupled protons of the cage organic groups for which the second term of Eq. (\ref{eqn2}) is estimated to be less than 5 kHz. For these protons the modulation frequencies match exactly 2$\nu_H$ and its higher harmonics, as can be seen from Fig.\ref{fig3}(b). On the other hand, a careful observation at the low-frequency part of spectra reveals that the first silicon peak is shifted by about 75 kHz from 2$\nu_{Si}$ whereas for the higher harmonics the shifts are - as expected - more pronounced. We therefore deduce that the silicon hyperfine coupling is comparable to $\nu_{Si}$, in line with the previously obtained parameters $|A|=$2.06 MHz and $|B|=$0.45 MHz for the inner silicon atom of the cage \cite{Mitrikas2012}.

In order to quantify these observations we performed numerical calculations using two different approaches: for $^{29}$Si we followed a complete simulation of the pulse sequence, whereas for the $^1$H nuclear spin bath the calculation was done considering the filter function of the CPMG sequence and a Gaussian function for the noise spectral density \cite{sup}. Fig.\ref{fig3}(a) and (b) show that both time and frequency domain data are in very good agreement with the simulated ones. For protons, where the free fitting parameters were the Gaussian function constants, the inferred noise spectral densities are centered at the $^1$H Larmor frequency, have a maximum of about 0.46 $\mu$T/$\sqrt{\mathrm{Hz}}$ and a broadening of 15 kHz (Fig.\ref{fig3}(c)). From these Gaussian functions we calculate an rms magnetic field noise $\sqrt{\int{{S_B(\omega)}^2d\omega}}=$0.09 mT which is in good agreement with the observed EPR linewidth of 0.1 mT.

\begin{figure}
%\centering
  \includegraphics{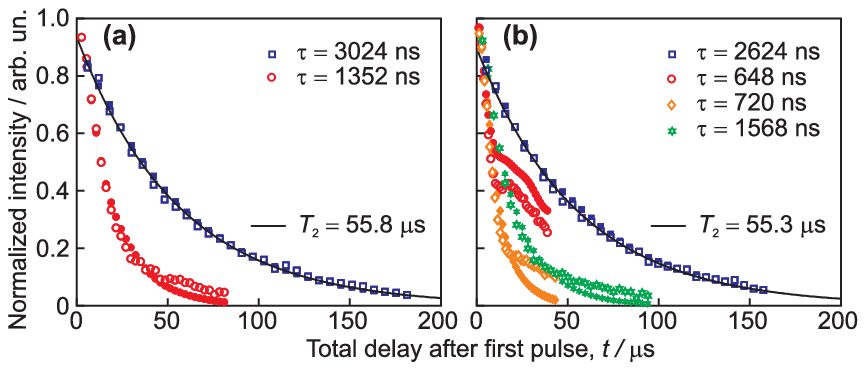}
  \caption{(Color online) CPMG decays measured with different $\tau$ values and $N=30$ pulses. Open symbols, experimental data; filled symbols, simulations. (a) High-field EPR transition, $B_0=$369.5 mT. (b) Low-filed EPR transition, $B_0=$318.6 mT. Black lines correspond to monoexponential fits with time constants $T_2$.}
  \label{fig4}
\end{figure}

The validity of our model was also tested by simulating CPMG decays for different $\tau$ values. Fig.\ref{fig4} shows that the simulations reproduce well the experimental data and this further supports the argument that it is the proton high-frequency noise that accounts for the fast decoherence under special conditions of dynamical decoupling. On the other hand, a proper choice of $\tau$ suppresses both low- and high-frequency noise leading to the longest time constant of about 56 $\mu$s (Fig.\ref{fig4}, squares). This value is smaller than $T_1$ and the expected $T_2$ for the electron spin concentration of the sample under study \cite{sup}. Moreover, it is virtually constant for proper $\tau$ values fulfilling $400$ ns$<\tau<4000$ ns, indicating a white-noise behavior for frequencies between 60 kHz and 600 kHz, similar to a "frozen-core" effect predicted for Si:P \cite{Sousa}. Future substitution of magnetic nuclei of the POSS cage (e.g. by perdeuteration) will elucidate whether this noise belongs to the tail of the proton low-frequency spectral density.

In conclusion, the application of the CPMG sequence on H@Q$_8$M$_8$ has shown that besides nuclear spin diffusion there are also two other sources of decoherence related to the nuclear spin bath: first, the weakly coupled protons of the organic groups of the cage behave like a fluctuating magnetic field with the proton Larmor frequency, $\nu_H$. This field induces a random phase on the central electron spin and creates additional dephasing. Second, for molecules possessing one or more $^{29}$Si nuclei in the core of the cage, the electron spin coherence is additionally modulated because the anisotropic hyperfine interaction is of the order of the nuclear Zeeman interaction and thus induces a degree of entanglement between the electron and the $^{29}$Si nuclear spin. These two sources of decoherence can be described as two different limits of the electron-nuclear hyperfine coupling \cite{Wrachtrup_NatNano_2012,Wrachtrup_PRL_2012} but are not manifested in a typical Hahn echo experiment. Our work shows that although dynamical decoupling with multi-pulse sequences can significantly suppress low-frequency noise, it may also enhance decoherence if the system bears a source of high-frequency noise; this should have high impact on quantum computing schemes based on dynamical decoupling and solid-state spin qubits. Moreover, the increase of the phase memory time up to $T_{\text{M}}$=56 $\mu$s in a chemically unmodified environment, poses atomic hydrogen encapsulated in POSS cages as a promising hybrid electron-nuclear spin system for quantum computing applications, comparable to endohedral fullerenes.

\end{document}

\begin{center}
\textbf{Supplemental Material}

for the manuscript

\textbf{Extending the Electron Spin Coherence Time of Atomic Hydrogen by Dynamical Decoupling}

by

George Mitrikas, Eleni K. Efthimiadou, and George Kordas

\emph{Institute of Advanced Materials, Physicochemical Processes, Nanotechnology and Microsystems, NCSR Demokritos, 15310 Athens, Greece}

email: mitrikas@ims.demokritos.gr
\end{center}

\section{\label{sec:level1} Theory and Simulations}
The numerical calculation of the CPMG signal can in principle be done using the spin Hamiltonian of the electron-nuclear coupled spin system under the unitary transformations of the pulse train. While this is straightforward for one $^{29}$Si nucleus, it becomes virtually impossible for $^1$H nuclear spins due to their large number and the wide distribution of their magnetic parameters. Therefore, for the latter case we assumed a Gaussian function for the noise spectral density of the $^1$H nuclear spin bath and the calculation of the CPMG decay was done by considering the filter function of the pulse train.
\subsection{\label{sec:level2} Echo modulations due to coupling with a single inner-cage $^{29}$Si spin}
The rotating frame spin Hamiltonian is given by \cite{Schweiger2001}
\begin{equation}
{\cal H}_0={\Omega}_SS_z+{\omega}_II_z+AS_zI_z+BS_zI_x,
\label{eq1}
\end{equation}
where $\Omega_S=\omega_S-\omega_{mw}$ is the offset of the electron Zeeman frequency $\omega_S=g\beta_e\textit{B}_0/\hbar$ from the mw frequency $\omega_{mw}$, $\omega_I=-g_n\beta_n\textit{B}_0/\hbar$ is the nuclear Zeeman frequency, $g$ and $g_n$ are the electron and nuclear g-factors, $\beta_e$ and $\beta_n$ are the Bohr and $^{29}$Si nuclear magnetons, $B_0$ is the static magnetic field along \textit{z}-axis, and \textit{A}, \textit{B} describe the secular and pseudo-secular part of the hyperfine coupling. For an axially symmetric hyperfine coupling tensor, [$A_{\bot}, A_{\bot}, A_{\|}$]=[$a_{\text{iso}}-T$, $a_{\text{iso}}-T$, $a_{\text{iso}}+2T$], where $a_{\text{iso}}$ and $T$ are the isotropic the anisotropic part of the hyperfine interaction, we have
\begin{equation}
A=a_{\text{iso}}+T(3\text{cos}^2{\theta}-1),\,\,\, B=3T\text{cos}{\theta}\,\text{sin}{\theta},
\label{eq2}
\end{equation}
with $\theta$ being the angle between the static magnetic field $B_0$ and the $z-$principal direction of the hyperfine tensor.
Starting from the equilibrium density operator $\sigma_{\text{eq}}=-S_z$, the first $(\pi/2)_x$ pulse creates electron spin coherence $\sigma_0=R^{-1}_x(\pi/2)\sigma_{\text{eq}}R_x(\pi/2)$, where $R_x(\pi/2)=\text{exp}[-i\pi/2S_x]$. After the $N'$th $(\pi)_y-$pulse the echo intensity is given by
\begin{equation}
\langle\sigma_x\rangle_{Si}=\text{Re Tr}(S_x\sigma_N),
\label{eq3}
\end{equation}
where $\sigma_N$ is the density matrix at the time of the $N'$th echo formation, $t=2N\tau$,
\begin{equation}
\sigma_N=U^{-1}_{\tau}R^{-1}_y(\pi)U^{-1}_{\tau}\sigma_{N-1}U_{\tau}{R_y}(\pi)U_{\tau},
\label{eq4}
\end{equation}
and $U_{\tau}=\text{exp}[-i{\cal H}_0\tau]$ is the propagator of free evolution.

\begin{figure}
\centering
  \includegraphics{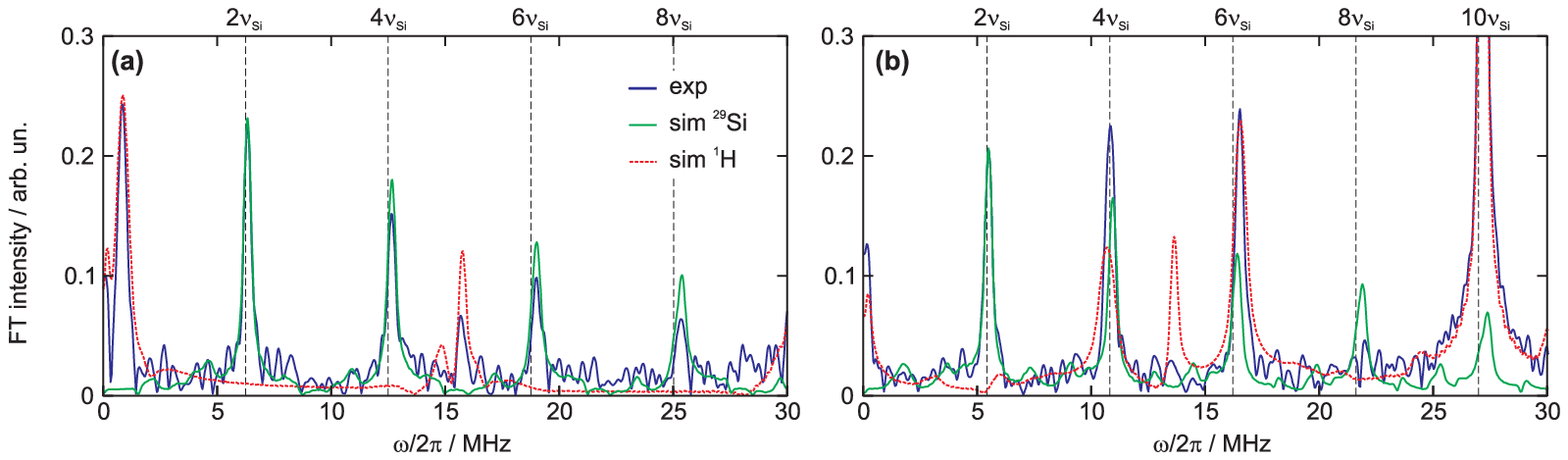}
  \caption{(Color online) Absolute value FT spectra of experimental (blue lines) and separately simulated (green-$^{29}$Si, red-$^1$H) signals of the last echo of a CMPG sequence with $N=$10. (a) High-field EPR transition, $B_0=$369.5 mT. (b) Low-filed EPR transition, $B_0=$318.6 mT. Black dashed lines denote even harmonics of $\nu_{Si}$.}
  \label{S1}
\end{figure}

From our previous HYSCORE study \cite{Mitrikas2012} we obtained $a_{\text{iso}}\simeq-$2 MHz and $T\simeq-$0.6 MHz for the inner $^{29}$Si cage atoms. We used these values as starting point for the simulations which included integration over all possible orientations. Refinement of the parameters gave $a_{\text{iso}}=-$2.2 MHz and $T=-$0.65 MHz that reproduced very well the position and intensity of peaks close to even harmonics of $\nu_{Si}$ (see Fig. \ref{S1}). Note that the high-field spectrum is more appropriate for this comparison because most of the $^{29}$Si peaks do not overlap with $^{1}$H harmonics.

\subsection{\label{sec:level3} Echo modulations due to coupling with the $^1$H spin bath}

\begin{figure}
\centering
  \includegraphics{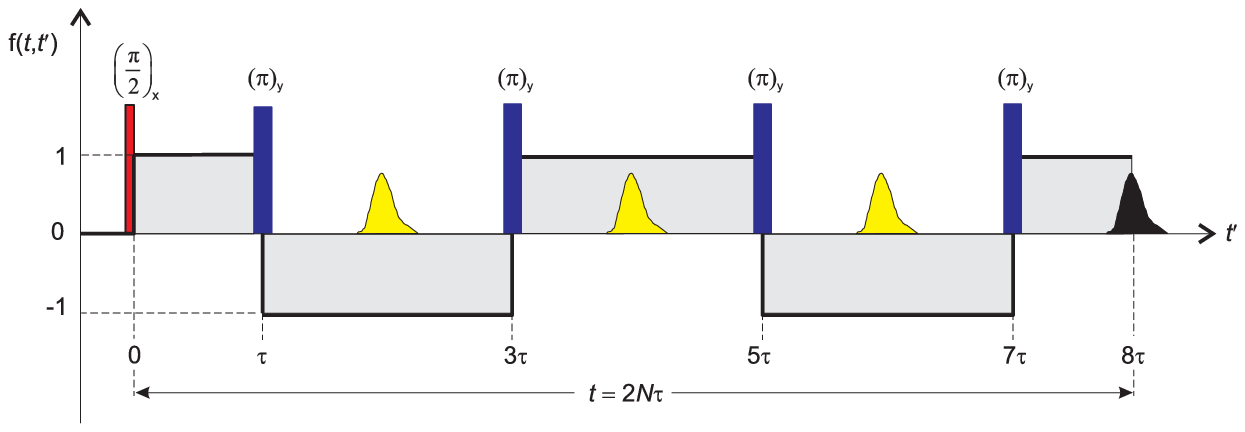}
  \caption{(Color online) CPMG sequence for $N=$4 and the corresponding function $f(t,t')$.}
  \label{S2}
\end{figure}

We treat the surrounding $^1$H nuclear spin bath as a source of classical noise with spectral density $S(\omega)$, i.e. a fluctuating magnetic field whose frequency is peaked at the $^1$H Larmor frequency $\nu_H$ and distributed according to $S(\omega)$. The coherence decay is given by \cite{Sarma}
\begin{equation}
\langle\sigma_x\rangle_H=\text{exp}(-\chi(t))\,\,\ \text{with} \,\,\ \chi(t)=\int_0^\infty\frac{d\omega}{2\pi}S(\omega)|\widetilde{f}(t;\omega)|^2,
\label{eq5}
\end{equation}
where $\widetilde{f}(t;\omega)$ is the filter function of the pulse sequence given by the Fourier transform of the function $f(t,t')$ with respect to $t'$,
\begin{equation}
\widetilde{f}(t;\omega)=\int_{-\infty}^\infty f(t,t')\text{e}^{-i\omega t'}dt'.
\label{eq6}
\end{equation}
For the CPMG sequence the function $f(t,t')$ is the Heaviside step function shown in Fig. \ref{S2} for $N=$4.
For the $^1$H spectral density we assumed a Gaussian distribution of width $\sigma$,
\begin{equation}
S(\omega)=\lambda^2\frac{1}{\sigma\sqrt{2\pi}}\text{exp}\left[ -\frac{(\omega-\omega_L)^2}{2\sigma^2}\right],
\label{eq7}
\end{equation}
where $\omega_L$ is the $^1$H Larmor frequency and $\lambda$ is the coupling strength parameter having units of angular frequency. Another useful representation of this noise is through the magnetic field sensitivity, $S_B(\omega)$, defined as
\begin{equation}
S_B(\omega)=\frac{2\pi}{\gamma_e}\sqrt{\frac{S(\omega)}{2\pi}},
\label{eq8}
\end{equation}
where $\gamma_e/2\pi=28.025\times10^9 \text{ Hz/T}$ is the gyromagnetic ratio and $S_B(\omega)$ has units of $\text{T/}\sqrt{\text{Hz}}$. Fig. 3(c) in main text shows noise spectral densities in this representation.

\begin{figure}
\centering
  \includegraphics{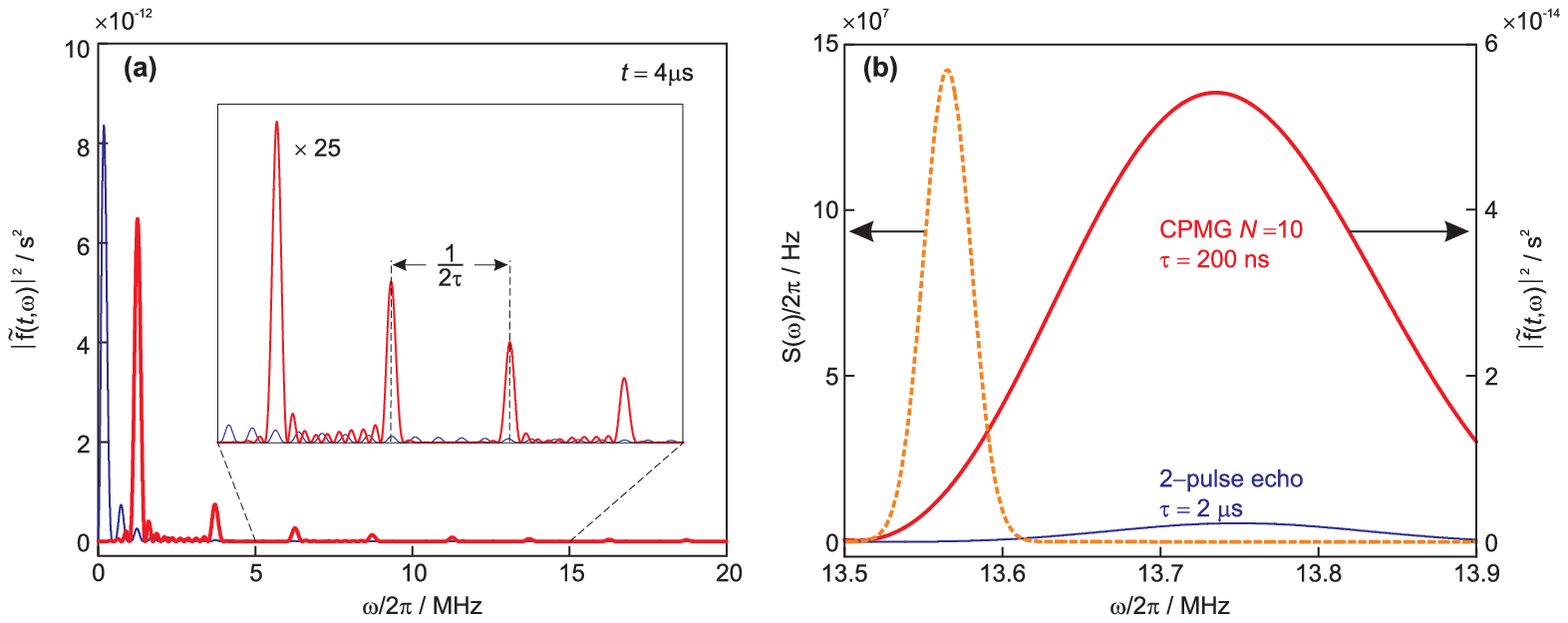}
  \caption{(Color online) (a) Comparison of the filter functions for $t=4 \mu$s between the two-pulse echo (blue thin line) and CPMG with $N=10$ (red thick line). (b) Details of the filter functions in the high frequency region together with the simulated $^1$H spectral density (orange dashed line) for the low-field EPR transition. }
  \label{S3}
\end{figure}

Fig. \ref{S3} compares the filter functions of the two-pulse echo and CPMG sequence with $N=10$ for the same time evolution $t=$4 $\mu$s. Fig. \ref{S3}(a) shows that, compared to two-pulse, the CPMG sequence not only suppresses low-frequency noise but also exhibits peaks at the high- frequency region where the $^1$H noise appears. When the filter and noise spectral density functions overlap (Fig. \ref{S3}(b)), additional decoherence occurs.
\subsection{\label{sec:level4} Overall simulation of echo modulations}
The total echo modulation is calculated as the product $\langle\sigma_x\rangle_{H}$$\langle\sigma_x\rangle_{Si}$ taking into account the natural abundance 4.68\% of $^{29}$Si. Considering only the Si$_8$O$_{12}$ core of the POSS cage, the binomial distribution predicts that $c_0=68.13\%$ of the cages have no $^{29}$Si atom. Therefore, these species contribute only to $^1$H modulations. The rest $c=31.87\%$ of the cages have one ($c_1=26.78\%$), two ($c_2=4.61\%$), three ($c_3=0.45\%$), four ($c_4=0.03\%$) etc. $^{29}$Si atom(s) in the core and therefore contribute also to $^{29}$Si modulations. Assuming that all eight $^{29}$Si atoms are magnetically equivalent, the total echo modulation is given by
\begin{equation}
\langle\sigma_x\rangle=\sum_{j=0}^{8} c_j\langle\sigma_x\rangle_{H}\left[\langle\sigma_x\rangle_{Si})\right]^j
\label{eq9}
\end{equation}
In our simulations we used only the first four terms of Eq. \ref{eq9} (up to $j=3$) because the rest of them have negligible contribution.
\begin{figure}[h]
\centering
  \includegraphics{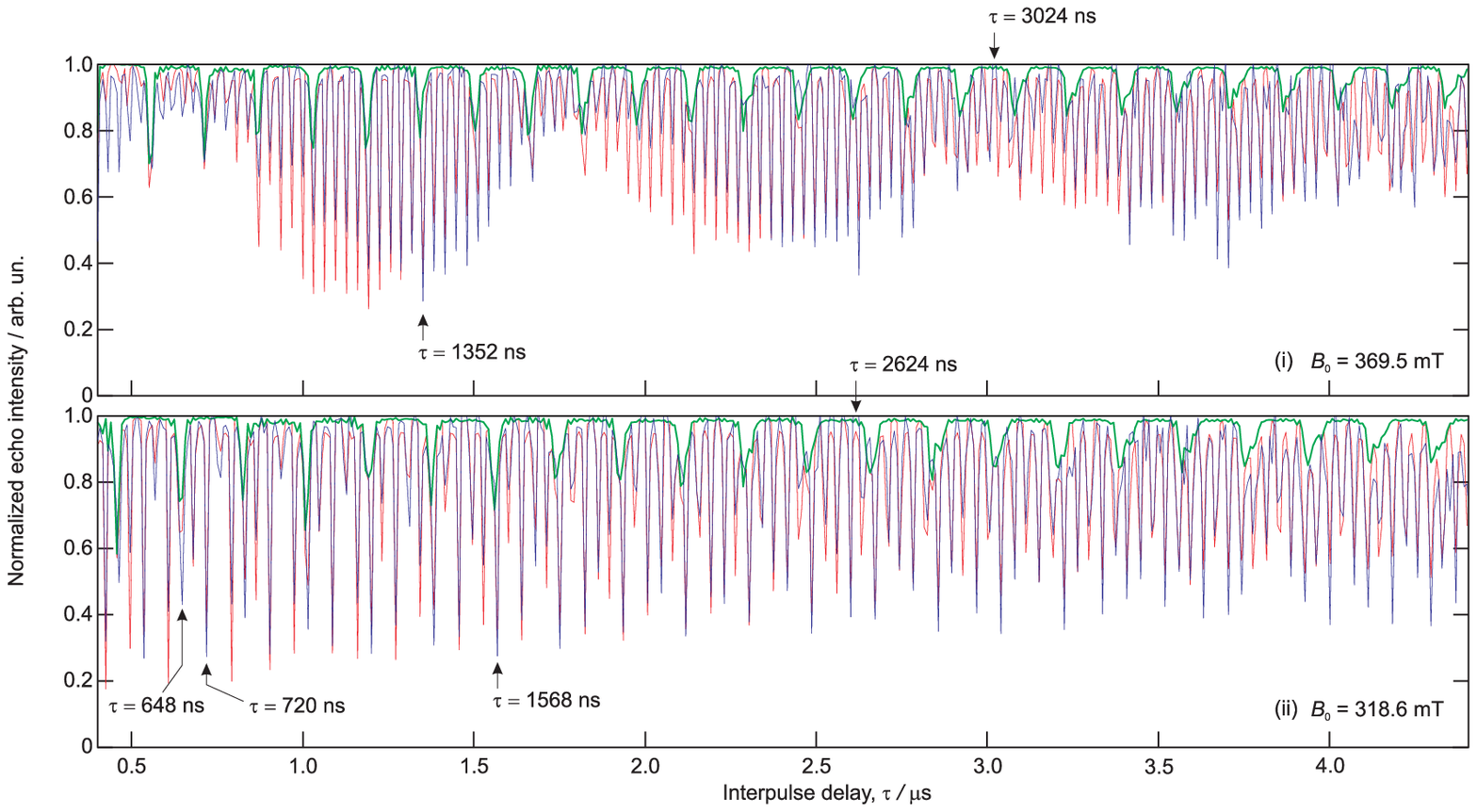}
  \caption{(Color online) Baseline-corrected CPMG decays obtained with $N=$10 (blue traces) and their numerical simulations (red traces) for the two EPR transitions. Green traces are the contribution of $^{29}$Si modulations to the total signal. Arrows mark the $\tau$ values used in measurements shown in Fig.4}
  \label{S4}
\end{figure}
Fig. \ref{S4} (green thick lines) shows the contribution of $^{29}\text{Si}$ modulations to the total signal along with the experimental data. Setting $\lambda$ and $\sigma$ as free varying parameters, the noise spectral densities were obtained by minimizing the sum of the residuals. The results gave $\lambda=1.53\pm0.05\times10^7$ rad/s and $\sigma/2\pi=14.8\pm0.5$ kHz for the high-field, and $\lambda=1.45\pm0.05\times10^7$ rad/s and $\sigma/2\pi=15.0\pm0.5$ kHz for the low-field EPR transitions. This broadening corresponds to static magnetic field $B_0$ inhomogeneity of about 0.3 mT which is much larger than the instrumental magnetic field inhomogeneity (using the same amount of sample the resolution of the liquid-state EPR spectrum acquired with the same spectrometer was about 10 $\mu$T). Therefore, we can assign this linewidth to inhomogeneous broadening of the proton nuclear spin spectrum. Moreover, the obtained $\sigma$ values are comparable to the dipolar couplings between methyl group protons, i.e. 20 kHz, showing that nuclear spin flip-flops driven by dipolar interactions are indeed quite probable.

\subsection{\label{sec:level5} Discussion on the obtained coherence time $T_2=56\text{ }\mu$s}
Here we discuss the observed $T_2$ value as a possible result of other relaxation effects like instantaneous diffusion (ID) and electron spin-spin interactions. The estimated electron spin concentration of our sample is about $C=1.3\times10^{16}$ spins/cm$^3$. Since the two EPR transitions are very well separated, the time constant due to ID is given by \cite{Schweiger2001}
\begin{equation}
\frac{1}{T_{\text{2(ID)}}}=C\frac{\pi}{18\sqrt{3}}\frac{\mu_0g^2\beta^2_e}{\hbar} \text{ sin}^2\frac{\theta_2}{2},
\label{eq10}
\end{equation}
where $\mu_0$ is the permeability of free space and $\theta_2$ the rotation angle of the refocusing pulse. Setting $\theta_2=\pi$ and $g=2$ gives $T_{2\text{(ID)}}=186 \text{ }\mu$s which is larger than both relaxation times $T_1=133 \text{ }\mu$s and $T_2=56 \text{ }\mu$s at this temperature ($T=200\text{ K}$). Moreover, similar measurements on a sample of half concentration revealed the same $T_2$. We can therefore safely conclude that ID is not the dominant relaxation mechanism that determines the CPMG decay. On the other hand, $T_1$-induced spectral diffusion is possible because due to the large hyperfine splitting we cannot excite both EPR transitions simultaneously. Although CPMG refocuses interactions between resonant (A) and non-resonant (B) electron spins, random flips of the B spins with relaxation time $T_1$ will contribute to the echo decay. This contribution is both temperature and concentration dependent. Preliminary measurements at very low temperatures ($T=5 \text{ K})$ showed that the spin-lattice relaxation time becomes extremely long, $T_1=57\pm2 \text{ s}$, however, the maximum observed CPMG time constant of $T_2\approx60 \text{ }\mu$s remained practically unchanged. This is a strong indication that $T_1$-spectral diffusion effects do not determine this decay.

Another mechanism that can contribute to decoherence is flip-flopping between neighboring electron spin pairs driven by dipole-dipole interactions. This includes transfer of magnetization between the central and a neighbor electron spin (spin diffusion) or change of the resonance frequency of the central electron spin due to flip-flops between two neighbors ($T_2$-type spectral diffusion). For this to be significant, the dipolar coupling of the two spins involved has to be larger or at least of the same order of magnitude as the difference in their resonance frequencies. The dipole-dipole coupling between two electron spins is given (in frequency units) by \cite{Schweiger2001}
\begin{equation}
\nu_{dd}=\frac{\mu_0}{4\pi h}\frac{g_1g_2\beta^2_e}{r^{3}_{12}}(1-3\text{cos}^2{\theta_{12}}),
\label{eq11}
\end{equation}
where $r_{12}$ is the distance between the two spins and $\theta_{12}$ is the angle between the static magnetic field $B_0$ and the inter-spin axis. Because the frequency difference of the two EPR transitions is very large (about 1.42 GHz), the effective concentration of the spins that can undergo flip-flops is $C/2\simeq6.5\times10^{15}$ spins/cm$^3$ that gives an average spin separation $r_{12}=65 \text{ nm}$. For this distance the average dipolar coupling frequency is about 90 Hz, much smaller than the inhomogeneous broadening of about 3 MHz of the EPR lines. We therefore conclude that flip-flopping mechanisms involving electron spins are highly suppressed in our system.

\section{\label{sec:level6} Materials and methods}

\subsection{\label{sec:level7} Sample preparation}

Octakis (trimethylsilyloxy) silsesquioxane (Si$_8$O$_{12}$(OSiMe$_3$)$_8$ or Q$_8$M$_8$, CAS 51777-38-9) was prepared following a two-step process:

Octaanion salt synthesis (\textbf{1}): The octaanion [Si${_8}$O$_{20}$$^{8-}$]$\cdot$8[Me$_4$N]$^+$ was prepared according to Hagiwara et al. \cite{D4R}. The mixture of TEOS/TMAOH/H$_2$O/EtOH in molar ratio 1/1/10/10 was left for vigorous stirring for 3 days. After this reaction time, the mixture concentrated by vacuum and cooled. Hydrated crystals of the octaanion were observed. These crystals were dehydrated after heating at 60 $^\circ$C in high pressure vacuum. 1g of the isolated product was dissolved in a mixture of pyridine/THF (150/100) which were pretreated in an ice bath. After 30 min stirring the unreacted products were removed under reduced pressure. A portion of HCl was added to neutralize the pyridine excess. The precipitated salt was isolated with filtration becoming a clear solution. The liquid was concentrated and the octaanion salt resulted.

[Si$_8$O$_{20}$][Si(CH$_3$)$_3$]$_8$ (\textbf{2}): Trimethylchlorosilane (22.0 ml, 0.189 mol) was added in hexane (120 ml) and the mixture was stirred for 30 min at 0 $^\circ$C. The flask equipped with a reflux condenser and an additional funnel aiming at adding the octaanion butanol solution dropwise (1.0 g octaanion in 50.0 ml butanol) under N$_2$ during 30 minutes period. The reaction was left for stirring additional 30 minutes and after that the mixture was extracted with hexane/brine twice and the organic layer was recovered. The organic layer was dried over Na$_2$SO$_4$. The hexane was concentrated to give a white solid in a good yield. Compound 2 was characterized by FT-IR and their characteristic peaks are: Si-O-Si at 1100 cm$^{-1}$, Si-CH$_3$ at 1260 cm$^{-1}$ \cite{Has2004}.

Encapsulation of atomic hydrogen was performed with $\gamma$-irradiation in steps using a $^{60}$Co source. The accumulated dose was measured with Red Perspex Dosimeters, Type 4034 AD. 54 mg of POSS powder were placed in a sealed vial filled with O$_2$ (acting as radical scavenger)  and irradiated for 20.9 days resulting in a total dose of 255 kGy. After $\gamma$-irradiation, 27.8 mg of the sample was transferred into an EPR quartz tube. Comparison of the cw EPR spectrum (in presence of O$_2$, unsaturated conditions) with a standard sample gave an estimate of $1.3\times10^{16}$ spins/cm$^3$ for the electron spin concentration. To avoid paramagnetic oxygen in pulsed EPR measurements, the quartz tube was filled with He gas following the procedure described previously \cite{Mitrikas2012}.

\subsection{\label{sec:level8} Pulsed EPR spectroscopy}
\begin{figure}[h]
\centering
  \includegraphics{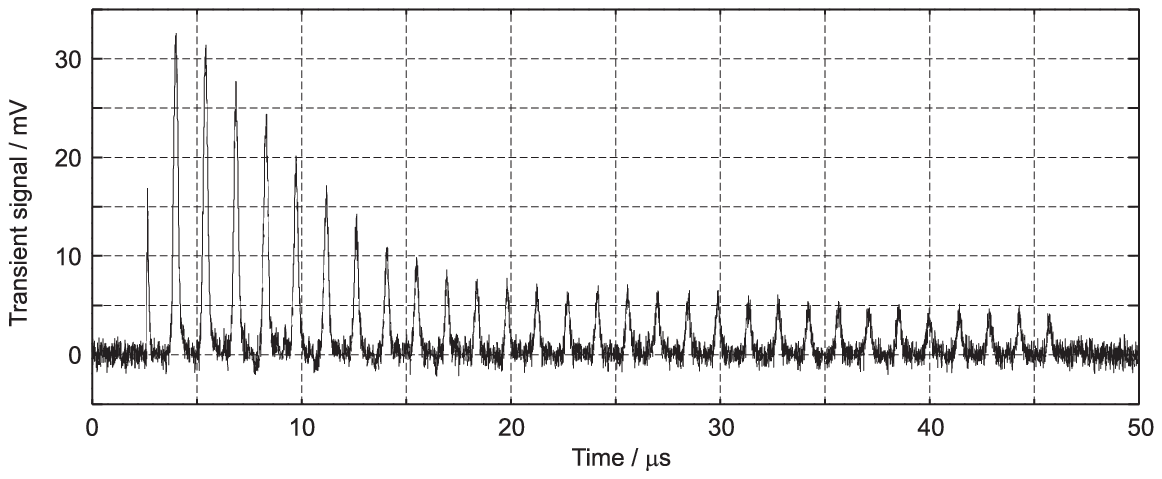}
  \caption{(Color online) Oscilloscope-recorded CPMG time trace for $N=30$ and $\tau=720\text{ ns}$. Number of averaging measurements, 1024; total acquisition time, 30 s. The trace is the difference between on- and off-resonance signals in order to remove unwanted baseline artifacts induced by microwave pulses.}
  \label{S5}
\end{figure}

EPR measurements at X-band ($\omega_{mw}/2\pi=9.698\text{ GHz}$) were carried out on a Bruker ESP 380E spectrometer equipped with an EN 4118X-MD4 Bruker resonator. Measurements at $T=200\text{ K}$ were performed with a helium cryostat from Oxford Inc. The microwave frequency was measured with a HP 5350B microwave frequency counter and the temperature was stabilized with an Oxford ITC4 temperature controller. Two-pulse echo and GPMG measurements used $t_{\pi/2}=16$ ns and $t_{\pi}=32$ ns. For the CPMG sequence the maximum number of pulses $N$ is limited by the spectrometer to 30. For CPMG experiments with constant $\tau$ values the measurements were recorded with a HP Infinium 54810A oscilloscope which allowed the acquisition of the entire time trace in a single shot (see Fig. \ref{S5}). All measurements were performed with repetition rates smaller than 1/5$T_1=1.5\text{ kHz}$ in order to avoid saturation.
\subsection{\label{sec:level9} Data manipulation}

The data were processed with the program MATLAB 7.0 (The MathWorks, Natick, MA). For $\tau$ values corresponding to maximum echo intensities in CPMG measurements, $T_2$ relaxation times were determined by fitting the time traces with single exponential functions, whereas for obtaining $T_M$ from two-pulse echo decay a stretched exponential function was appropriate. The time traces of variable $\tau$ CPMG experiments (Fig2. main text) were baseline corrected with a single exponential, apodized with a gaussian window and zero filled. After Fourier transform the absolute-value spectra were calculated. CPMG traces were simulated with a program written in-house based on the theory described above.